\documentclass[12pt]{article}

\usepackage[utf8]{inputenc}
\usepackage[T1]{fontenc}
\usepackage{authblk} %Pour affiliatiaon
\usepackage{amsmath,amsfonts,amssymb,amsthm}
\numberwithin{equation}{section}
\usepackage{dsfont} %Pour math ds
\usepackage{color}
\usepackage[dvipsnames]{xcolor} % for more colours and to invent some
\colorlet{mygreen}{green!60!black}
\colorlet{mygrey}{white!60!black}
\usepackage{graphicx} % pour l'inclusion de figures en eps,pdf,jpg
\usepackage{float}  %Permet de placer la figure ou tu veux
\usepackage{booktabs} %Permet \toprule et \midrule dans l'env. tabular
\usepackage{setspace} %Pour changer l'interligne
\usepackage{multicol} %Pour des équations a plusieurs col.
\usepackage{url} 
\usepackage[dvipsnames]{xcolor}
\usepackage[
  colorlinks = true,
  linkcolor  = MidnightBlue,  % equations, sections, figures
  citecolor  = MidnightBlue,  % bibliography citations
  urlcolor   = MidnightBlue   % URLs, DOIs
]{hyperref}

\usepackage[numbers,sort&compress]{natbib}

\usepackage{tikz}
\usetikzlibrary{shapes,positioning,decorations.pathreplacing} % semble necessaire pour qcircuit
\usetikzlibrary{calc} % pour utiliser la coordonnée d'un noeud
\usetikzlibrary{arrows.meta,decorations.pathmorphing,shapes.geometric,decorations.markings,intersections}% pour les tailles de têtes de fleches & autres commandes pour spacetime diagram

\newlength{\wirespace} % in the usual diagram
\newlength{\wiresmall} % in the phone, big spacing
\newlength{\boxsize}
\usepackage{geometry} %Mise en page
\usepackage[font=footnotesize,labelfont=bf]{caption} %Permet de changer le font size des caption
\usepackage{palatino} %Fonts

\def \be {\begin{equation}} 
\def \ee {\end{equation}}
\def \bes {\begin{equation*}}
\def \ees {\end{equation*}}
\def \baa {\begin{align}}
\def \eaa {\end{align}}
\def \baas {\begin{align*}}
\def \eaas {\end{align*}}
\def \bea {\begin{eqnarray}}
\def \eea {\end{eqnarray}}
\def \beas {\begin{eqnarray*}}
\def \eeas {\end{eqnarray*}}

\newcommand{\ket}[1]{\rvert#1\rangle}
\newcommand{\bra}[1]{\langle #1\rvert}

\newcommand{\hil}{\mathcal{H}^}
\newcommand{\ketbra}[2]{| #1 \rangle \langle #2 |}

\newcommand{\deq}{\stackrel{\text{\tiny{def}}}{=}}

\newcommand{\bol}{\boldsymbol} %boldsymbol

\newcommand{\Ry}[1]{R_{#1}} %y Rotation
\newcommand{\cc}[1]{c_{#1}} % cos
\newcommand{\s}[1]{s_{#1}} % sin
\newcommand{\Uf}{\mathsf U} % Functional representation
\newcommand{\cnot}{{\text{Cnot}}}
\newcommand{\ctrlu}{\text{Ctrl-$U$}}

\newcommand{\plustwo}{+2}
\newcommand{\cplusone}{\text{C}_{+1}}
\newcommand{\plusone}{+1}
\newcommand{\plusthree}{+3}
\newcommand{\plusk}{+k}
\newcommand{\notgate}{\text{not}}

\newcommand{\one}{\mathds{1}}
\newcommand{\whole}{\mathfrak U} % Whole system
\newcommand{\sqb}[1]{\boldsymbol s^{_{(#1)}}} %Decripotrs s with high t 
\newcommand{\qt}[2]{q_{#1}(#2)} %Normal notation
\newcommand{\qbt}[2]{\boldsymbol q_{#1}(#2)}
\title{Explaining Bell Locally}

\author{Charles Alexandre B\'edard}
\affil{\small {École de technologie supérieure}\\
\footnotesize \emph{charles.alexandre.bedard@etsmtl.ca}}
\date{November 2025}

\begin{document}
\maketitle
\thispagestyle{empty} %Removes page number

%Nature requirements: 2500 mots, 4 figures.
% 270 mots par figure

\begin{abstract}
\noindent
In the Heisenberg picture of unitary quantum theory, Bell inequalities are violated with local elements of reality interacting locally. Here is how: Upon measuring her particle of the entangled pair, Alice—like other coupled systems—smoothly and locally evolves into two non-interacting versions of herself, each of which records a different outcome: she \emph{foliates}. Everything that suitably interacts with the Alices foliates in turn, generating worlds which, for all practical purposes, remain distinct and autonomous. At spacelike separation, an analogous yet independent process occurs to Bob when he measures his particle, locally differentiating him and his surroundings into two non-interacting instances. To confirm the violation of Bell inequalities, Alice and Bob must further interact to produce a record of the joint outcomes. The record arises from the two local worlds of Alice, and those of Bob, and foliates into four instances: ‘00’, ‘01’, ‘10’ and ‘11’. The outcomes that win the Clauser–Horne–Shimony–Holt (CHSH) game sum to a measure of~$\cos^2(\pi/8)$.
\end{abstract}

\section{Local Realism Beyond Local Hidden Variables}

In the century-long history of quantum theory, Bell’s theorem \cite{bell1964} stands out among the most thought-provoking results of its foundations. It reveals a conflict between quantum theory’s predictions and those allowed by so-called local hidden variable theories. Experimental vindications \cite{freedman1972experimental,aspect1981experimental,aspect1982experimental,aspect1982exp2,weihs1998violation,pan2000experimental,hensen2015loophole,giustina2015significant,shalm2015strong} of the quantum predictions were of Nobel Prize merit \cite{Nobelcomittee2022,brassard2023profile} and contributed to the prevalent conclusion that nature is non-local. In stark contrast with this orthodoxy, I show that the Heisenberg picture of unitary quantum theory, as developed by Deutsch \& Hayden \cite{deutsch2000information}, depicts an entirely local account.

How can remote measurements on a pair of entangled particles be both genuinely random yet perfectly correlated? That was the puzzle of Einstein, Podolsky and Rosen (EPR), who argued \cite{einstein1935can} that if locality is to be preserved and quantum theory is empirically adequate, then the theory is incomplete. A natural completion is given by \emph{hidden variables} thought to underpin the randomness of measurements while also explaining correlations. In spite of being unknown, or even perhaps unknowable, these underlying parameters would be determinative of the outcomes. A \emph{local} hidden variable theory is one in which all influences, including those by and on hidden variables, are constrained by the structure of spacetime, and hence limited by the speed of light. For instance, if two laboratories are spacelike separated, Alice’s freely chosen experimental setting neither affects Bob’s laboratory nor the hidden variables in its vicinity.

Building upon the EPR argument, Bell \cite{bell1964} established that quantum theory’s predictions are incompatible with an underlying description in terms of local hidden variables (see appendix~A for a basic explanation of Bell’s theorem via the Clauser–Horne–Shimony–Holt (CHSH) game). The incompatibility becomes manifest \cite{CHSH} in the experimental measurement of a quantity which ought to be smaller than a fixed bound if nature operates by local hidden variables. This is a \emph{Bell inequality}. According to quantum theory, however, Bell inequalities can be violated. And they are: Ever more convincing experiments systematically corroborate the quantum predictions, violating Bell inequalities and, thereby, refuting the local hidden variable account that Bell formalized.\footnote{In the literature on Bell’s theorem, ‘quantum non-locality’ is often used as a synonym for ‘incompatibility with local hidden variables’ or ‘incompatibility with Bell-locality’. However, this is a misleading linguistic shortcut, for the incompatibility with Bell-locality does not imply that the only valid explanations are fundamentally non-local.}

The most vocal conclusions assert \emph{non-locality}. Aspect declares that ‘we should renounce local realism’ \cite{aspect2015closing}, while according to Maudlin, ‘the physical world itself is non-local’ \cite{maudlin2014bell}. Some instead posit \emph{retrocausality}, which ‘includes the existence of backwards-in-time influences’ \cite{sutherland2008causally}, while others invoke \emph{superdeterminism}, the assumption that ‘the prepared state of an experiment is never independent of the detector settings’ \cite{hossenfelder2020rethinking}. But all of these tentative explanations are cast in a worldview shaped by hidden variables as if they were the only path to reconcile quantum theory with \emph{realism}, that is, the conjectured existence of a real physical world. Hidden variables were indeed a realist defence against the void enforced by the Copenhagen school, which effectively forbade inquiry into actual states of affairs between measurements. But by no means are they the sole realist program.

\begin{figure}[t]
    \centering
    \includegraphics[width=0.85\linewidth,trim=240 160 240 270,clip]{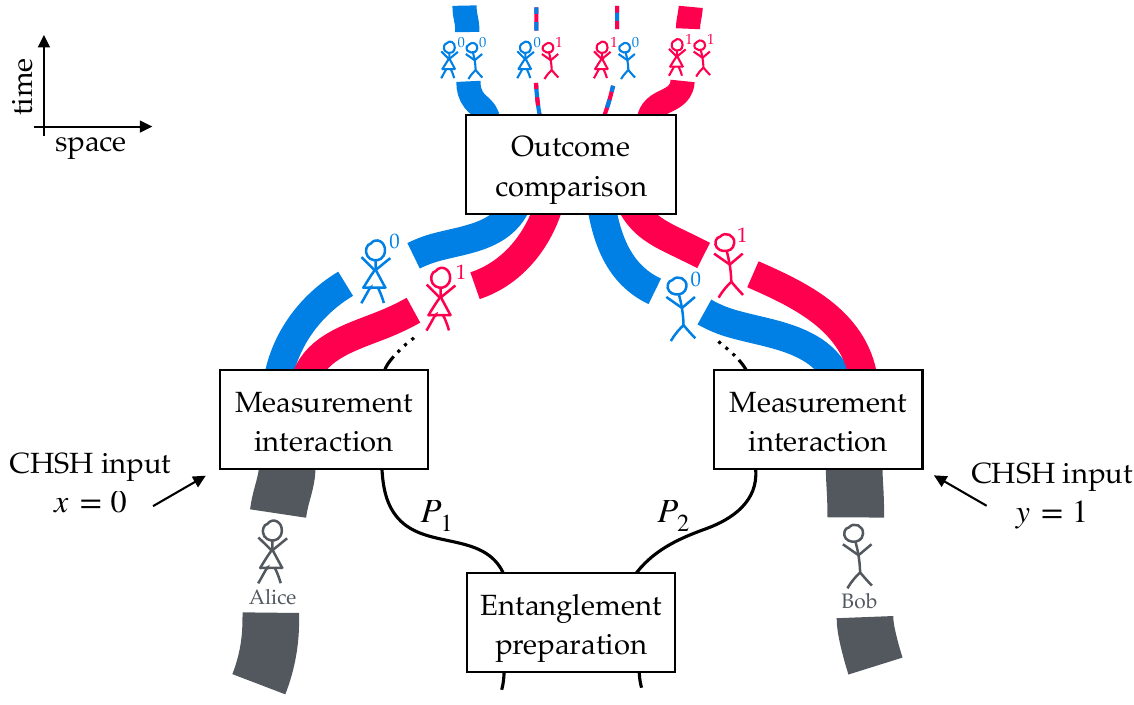}
    \caption{Schematic depiction of a Bell-inequality violation in the Heisenberg picture of unitary quantum mechanics. The
thickness of a branch reflects its probability measure.}
    \label{fig:diag}
\end{figure}

In a seemingly fantastic account reminiscent of the one outlined in the abstract and schematically depicted in figure~\ref{fig:diag}, Bruce \cite[p.~130--132]{bruce2004schrodinger} imagined that measurements in the Deutsch–Hayden formalism generate many local booths within which each observer obtains different outcomes. The local booths of the various Alices and those of the various Bobs pair into consistent Bell-violating classical worlds only when they compare results, via subluminal influences. As Raymond-Robichaud \cite{brassard2019parallel} points out, such accounts are manifestly local-realistic, and therefore contradict the widespread yet misconceived equivalence between local hidden variables and local realism. Within the local-fluids-in-spacetime framework—a supplement to quantum theory that adds further ontological assumptions—Waegell presents a similar account of Bell-like violations in terms of local entities \cite[\S5]{waegell2023local}.

As I shall show, this seemingly fantastic story of Bell violations is not merely a proof of principle in a toy model, nor does it demand an extension or a newly formed theory. \emph{It is how Bell inequalities are violated}, insofar as we rely on quantum theory, which I shall treat fully unitarily and in the Heisenberg picture as Deutsch~\& Hayden do.

\section{Locality from Unitarity in the Heisenberg Picture} \label{sec.progress}

Textbook quantum theory instructs a blatant inconsistency between two types of dynamical laws. On the one hand, all dynamical processes except `measurements' abide by unitary evolution, which is continuous, local, deterministic and in principle reversible. On the other hand, the so-called `measurements' follow a collapse rule that is discontinuous, non-local, stochastic and irreversible. Everett~\cite{everett1973theory} solved this inconsistency by dismissing the need for the ambiguous yet supposedly consequential notion of measurement. Measurements are, Everett argues, mere interactions. And observers, other quantum systems. He thus investigated measurement interactions within the unitary quantum theory and showed that in spite of being in superposition, observers do not directly witness their own superposition. Instead, they evolve alongside and in relation with other systems in independent and autonomous terms of the wave function, each describing a single classical world. Everett’s proposal not only saved the compatibility of quantum theory with scientific realism, but it made important progress on the question of locality, as it dismissed the instantaneous collapse of the wave function. As a remarkable side-effect of his solution, the totality of what we see around us, the universe, is but a tiny sliver of a much richer multiverse.

Adopting the universality of unitary evolution has provided key insights on the analyses of the EPR~\cite{page1982einstein} and Bell~\cite{tipler2014quantum,brown2016bell,waegell2020reformulating} scenarios, like the explanatory possibilities of observers in superposition and the crucial interaction in which observers compare their results, which must also be treated quantum-mechanically. Alongside these insights, many have pointed out, as Vaidman~\cite{VaidmanMWI} does, that Bell’s theorem ‘cannot get off the ground in the framework of [Everett] because it requires a single outcome of a quantum experiment’. While true, this observation is often treated as an endpoint: the coexisting outcomes are an excuse to avoid explaining Bell, rather than an opportunity to do so. ‘[I]n Everettian quantum mechanics, violations of Bell’s inequality are relatively uninteresting’, Wallace writes~\cite[ch.~8]{wallace2012emergent}. This disengagement may be due to the impossibility of extracting from the Schrödinger picture a fully local account. Only so much of locality can be saved in the Schrödinger picture: ‘in Everettian quantum mechanics interactions are local but states are non-local’~\cite[ch.~8]{wallace2012emergent}.

Einstein, whose steadfast defence of locality led to no less than the general theory of relativity, insisted that ‘the real factual situation of system~$S_2$ is independent of what is done with the system~$S_1$, which is spatially separated from the former’~\cite{schilppalbert1970}. This criterion of locality cannot be met by the modes of description offered in the Schrödinger picture. Indeed, if the wave function is intended to capture the real factual situation, then entanglement undermines the required independence: an action on system~$S_1$ alters the same wave function that also describes system~$S_2$. On the other hand, while reduced density matrices do maintain this independence, they fail to fully represent the real factual situation, as they provide an overly superficial account that omits how entangled systems can further interact.

Remarkably, Deutsch \& Hayden~\cite{deutsch2000information} demonstrated that the Heisenberg picture admits a mode of description that fulfils Einstein’s criterion. In this picture each system possesses a \emph{descriptor} that remains unchanged by actions on other systems and, furthermore, the descriptor of a composite system is nothing but the collection of individual descriptors pertaining to the subsystems. Crucially, these provide sufficient information to compute the distribution of any measurement performed on the composite system. In other words, the density matrix of the composite system~$S_1S_2$ can be obtained from the local descriptor of~$S_1$ and that of~$S_2$. Therefore, this underlying ontology is local in the double sense of having no action at a distance, and being separable—notwithstanding entanglement.

Descriptors have been used to solve important conundrums: superdense coding~\cite{bennett1992communication} and quantum teleportation~\cite{Bennett1993teleporting} have been explained completely locally~\cite{bedard2021abc,deutsch2000information,bedard2023teleportation}; the Aharonov–Bohm~\cite{aharonov1959significance} phase was shown to be locally acquired~\cite{marletto2020aharonov}; and a local-realistic treatment of indistinguishable particles led to the discovery of creation operators for non-abelian anyons~\cite{tibau2023locality,vidal2023creation}. Regarding Bell experiments, the pioneers of descriptors demonstrated~\cite{deutsch2000information} that the information about the rotation angles is fully localized, and by analysing the dynamics of specific observables, Rubin~\cite{rubin2001locality} showcased how observers evolve into labelled local copies. For descriptors, this amounts to their foliation into relative descriptors, a formalization by Kuypers~\& Deutsch~\cite{kuypers2021everettian} that is central to my analysis.

\section{Descriptors}

In this section, I provide an overview of how descriptors work. For a complete and pedagogical exposition, see Ref.~\cite{bedard2021abc}.

In the Heisenberg picture, observables evolve in time while the state is stationary and conveniently set to~$\ket{0}^{\otimes n}$ in quantum computational networks.\footnote{The implicit identification between the~$\ket{\hspace{-1pt}+1}$ and~$\ket{\hspace{-1pt}-1}$ eigenstates of the $z$~observable,~$\sigma_z$, and the~$\ket{0}$ and~$\ket{1}$ eigenstates of the computational observable $(\one - \sigma_z)/2$ is assumed.} The Heisenberg `state' is thus a misnomer; I shall refer to it as the reference vector. Instead, the object describing physical systems as they evolve must be tied to observables. But which observables should we track over time—should we attempt to track all of them? The infinitely many time-evolving observables can all be obtained from a \emph{generating set}, namely a set of operators that multiplicatively generate an operator basis. These operators can be chosen in a way that each acts non-trivially on one single system. For each given system, the generators acting on it are encompassed into the \emph{descriptor} of that system.

Let~$\whole$ be the whole system considered, and let~$\mathfrak Q_i$ denote the $i$\textsuperscript{th} qubit. At time~$0$ the descriptor of~$\mathfrak Q_i$ can be expressed as the pair of operators acting on~$\hil{\whole} \simeq \hil{\mathfrak Q_i} \otimes \hil{\overline{\mathfrak Q_i}}$,
\be \label{eq:initialdescriptor}
\qbt{i}{0} = \left(\qt{ix}{0},\, \qt{iz}{0}\right)
= \left(\sigma_x \otimes \one^{\overline{\mathfrak Q_i}},\, \sigma_z \otimes \one^{\overline{\mathfrak Q_i}}\right),
\ee
where~$\sigma_x$ and~$\sigma_z$ are the Pauli matrices, and~$\one^{\overline{\mathfrak Q_i}}$ is the identity operator on all but~$\mathfrak Q_i$. As they evolve, descriptors preserve their algebraic relations, so in particular, at any given time, the descriptor components of different qubits commute.

If between time~$t$ and~$t\!+\!1$ a gate~$G$ is applied to a set of systems labelled by~$\mathcal J$, the descriptor of qubit~$i$ evolves according to
\be \label{eq:step}
\qbt{i}{t+1}
= \Uf_G^{\dagger}\!\left(\left\{\qbt{j}{t}\right\}_{j \in \mathcal J}\right)\,
  \qbt{i}{t}\,
  \Uf_G\!\left(\left\{\qbt{j}{t}\right\}_{j \in \mathcal J}\right),
\ee
where~$\Uf_G(\cdot)$ is a fixed operator-valued function characteristic of the gate~$G$. Note that if~$i \notin \mathcal J$, the commutation of the components of~$\bol q_i$ with those of~$\bol q_j$ ensures that the descriptor of~$\mathfrak Q_i$ `is independent of what is done with the system' labelled by~$\mathcal J$, as mandated by Einsteinian locality.

Gates have an action on the systems they affect. The Hadamard gate acts as
\be \label{eqH}
H : (q_x, q_z) \to (q_z, q_x),
\ee
namely, it switches the components of the descriptor of the affected system, regardless of their expressions when acted upon. A Bloch-sphere rotation around the~$y$-axis transforms the descriptor as
\be \label{eqR}
\Ry{\theta} : (q_x, q_z) \to \left(\cos\theta\, q_x + \sin\theta\, q_z,\; -\sin\theta\, q_x + \cos\theta\, q_z\right).
\ee
And the action of the~$\cnot$ is
\be \label{eqCnot}
\cnot :
\left\{
\begin{array}{l}
  (q_{cx},\, q_{cz}) \\
  (q_{tx},\, q_{tz})
\end{array}
\right\}
\to
\left\{
\begin{array}{l}
  (q_{cx} q_{tx},\, q_{cz}) \\
  (q_{tx},\, q_{cz} q_{tz})
\end{array}
\right\},
\ee
where the label~$c$ refers to the control qubit and the label~$t$ to the target qubit.

Any observable~$\mathcal O(t)$ pertaining to a set of systems labelled by~$\mathcal J$ can be expressed as a polynomial in~$\{\bol q_j(t)\}_{j \in \mathcal J}$. Its expectation value~$\langle \mathcal O(t) \rangle$ is calculated by bracketing the reference vector~$\ket{0}^{\otimes n}$. Multiversal \emph{measures}, or probabilities, are obtained by such calculations. An observable~$\mathcal O(t)$ is \emph{sharp}~\cite{kuypers2021everettian} if, with respect to the Heisenberg state, it has a definite value, i.e. null variance, \mbox{$\langle\mathcal O(t)^2\rangle = \langle\mathcal O(t)\rangle^2$}.

Let~$\mathfrak S$ be a system with descriptor~$\bol s$ which at time~$t$ has a sharp observable. Let~$U$ be a unitary on~$\mathfrak S$ that transforms it into another value of the same sharp observable, and let~$\mathfrak Q_1$ be a control qubit with an unsharp~$\qt{1z}{t}$. The functional form of the controlled unitary is
\be \label{eq:ctrlu}
\Uf_{\ctrlu}(\bol q_1,\bol s)
= P_{+1}(q_{1z})\,\one
  + P_{-1}(q_{1z})\,\Uf_U(\bol s),
\ee
where~$P_{\pm1}(q_{1z}) \deq (\one \pm q_{1z})/2$ is a projector. From equations~\eqref{eq:step} and~\eqref{eq:ctrlu},
\bes
\bol s(t+1)
= \underbrace{P_{+1}(q_{1z}(t))\,\bol s(t)}_{\bol s_{+1}(t+1)}
+ \underbrace{P_{-1}(q_{1z}(t))\,
  \Uf_U^{\dagger}(\bol s(t))\,
  \bol s(t)\,
  \Uf_U(\bol s(t))}_{\bol s_{-1}(t+1)}.
\ees
This \emph{foliates}~\cite{kuypers2021everettian} the system~$\mathfrak S$ into two instances, the \emph{relative descriptors} $\bol s_{\pm1}(t+1) \deq P_{\pm1}(\bol q_{1z}(t))\,\bol s(t+1)$, each admitting a distinct value of a sharp observable. Under many forthcoming dynamics, they remain distinct and autonomous. For instance, any follow-up unitary~$V$ on~$\mathfrak S$ alone results in
\bes \label{eq:autonomy}
\sqb{t+2}
= \underbrace{\Uf_V^{\dagger}(\sqb{t+1}_{+1})\,
                \sqb{t+1}_{+1}\,
                \Uf_V(\sqb{t+1}_{+1})}_{\sqb{t+2}_{+1}}
 + \underbrace{\Uf_V^{\dagger}(\sqb{t+1}_{-1})\,
                \sqb{t+1}_{-1}\,
                \Uf_V(\sqb{t+1}_{-1})}_{\sqb{t+2}_{-1}}.
\ees

In the Schrödinger picture, the universe is described by a time-evolving wave function. In the Heisenberg picture, by contrast, it is described by time-evolving local generators of the algebra of observables. As has been shown previously~\cite{timpson2005nonlocality,wallacetimpson07,raymond2021local,bedard2021cost}, these two structures are not isomorphic: infinitely many sets of descriptors are consistent with any given wave function. This non-isomorphism implies that the two descriptions are not merely alternative presentations of the same underlying structure; rather, Heisenberg descriptors are formally richer, and they surject onto the Schrödinger wave function. The stories depicted in each picture cannot be put into one-to-one correspondence. For more on the non-isomorphism, and how it contrasts with the usual proof of equivalence between the two pictures, see Ref.~\cite{bedard2025Realism}.

\section{A Bell Experiment}\label{sec:exp}

The computational network of a Bell experiment is displayed in figure~\ref{fig:bell}. 
Particle~$1$, Particle~$2$, Alice and Bob are represented by qubits $\mathfrak Q_1$, $\mathfrak Q_2$, $\mathfrak Q_A$ and~
$\mathfrak Q_B$, respectively. 
Their initial descriptors are all of the form given by equation~\eqref{eq:initialdescriptor}, and because the Heisenberg reference vector is fixed to~$\ket 0$ on all systems, the qubits are initially sharp with their $z$ observables holding value~$+1$.
Charlie is represented by the 4-level system~$\mathfrak S_C$ corresponding to a thicker line on the network. 
It admits a computational observable spectrally decomposed as $\sum_{j=0}^{3} j \ketbra jj$, which is initially sharp in the~$0$\textsuperscript{th} eigenvalue.
Many possible generating sets of operators can used~\cite{raymond2021local, tibau2023locality} for the descriptor~$\bol s_C$, yet the choice is irrelevant for the current purposes, as I shall carry out the calculations without explicit dependence on their form.
The particles are entangled and then shared among Alice and Bob, who perform a rotation in accordance with the input of the Bell test.
Upon measuring their particles---measurements amount to \cnot s---Alice and Bob communicate their outcomes to Charlie, who records them.
The gate~`$+ k$' alters~$\ket j$ to~$\ket{j + k \text{~(mod $4$)}}$.

\begin{figure}
	\centering
	\resizebox{\linewidth}{!}{
	\begin{tikzpicture} [scale=1,  line width=0.75] 
		
		%beginning of wire
		\node (debut2) at (0,0){};
		\node (debut1) at ([yshift=\wirespace]debut2){};
		
		%initial descriptor
		\node[above right = -5pt and 0pt of debut1] (q10) {\scriptsize{$(q_{1x} \hspace{1pt},\, q_{1z})$}};
		\node[above right = -5pt and 0pt of debut2] (q20) {\scriptsize{$(q_{2x} \hspace{1pt},\, q_{2z})$}};
		
		%Hadamard
		\node[draw, rectangle, minimum width=\boxsize, minimum height=\boxsize, inner sep = 0, right = 44pt of debut1] (H) {$H$};
		
		%descriptor1
		\node[right = 18pt of q10] (q11) {\scriptsize{$(q_{1z} \hspace{1pt},\, q_{1x})$}};
		
		%Cnot1
		\node[draw, circle, fill=black, minimum width=0.14cm, inner sep=0pt, right= 42pt of H] (C1) {}; % Control
		\draw let  
    			\p1 = (C1) in
      			(\x1,\y1-\wirespace) circle (0.33\boxsize)
			node[inner sep = 0] (T1){}; %Target
		\draw let 
			\p1 = (T1) in
			(C1) -- (\x1, \y1-0.33\boxsize); %Link
			
		%descriptor2
		\node[right = 5pt of q11] (q12) {\scriptsize{$(q_{1z}q_{2x} \hspace{1pt},\, q_{1x})$}};
		\node[right = 69pt of q20] (q22) {\scriptsize{$(q_{2x} \hspace{1pt}, \, q_{2z}q_{1x})$}};

		%beginning of A and B
		\node (debutA) at ([xshift=290pt, yshift=\wirespace]debut1){};
		\node (debutB) at ([xshift=290pt, yshift=-\wirespace]debut2){};
		
		%initial descriptor of A and B
		\node[above right = -5pt and 0pt of debutA] (qA0) {\scriptsize{$(q_{Ax} \,, \, q_{Az})$}};
		\node[above right = -5pt and 0pt of debutB] (qB0) {\scriptsize{$(q_{Bx} \,, \, q_{Bz})$}};
		
		%Rotations	
		\node[draw, rectangle, minimum width=\boxsize, minimum height=\boxsize, inner sep = 0, right = 60pt of C1] (Rth) {$\Ry{\theta}$};
		\node[draw, rectangle, minimum width=\boxsize, minimum height=\boxsize, inner sep = 0, right = 61.4pt of T1] (Rph) {$\Ry{\phi}$};

		%descriptor3
		\node[right = 20pt of q12] (q13) {\scriptsize{$( \cc{\theta} q_{1z}q_{2x} + \s{\theta} q_{1x}  \hspace{1pt}, \, -\s{\theta} q_{1z}q_{2x} + \cc{\theta} q_{1x} )$}};
		\node[right = 20pt of q22] (q23) {\scriptsize{$( \cc{\phi} q_{2x} + \s{\phi} q_{2z}q_{1x}  \hspace{1pt}, \, -\s{\phi} q_{2x} + \cc{\phi} q_{2z}q_{1x} )$}};

		%CnotA
		\node[draw, circle, fill=black, minimum width=0.14cm, inner sep=0pt, right= 151pt of Rth] (CA) {}; % Control
		\draw let  
    			\p1 = (CA) in
      			(\x1,\y1+\wirespace) circle (0.33\boxsize)
			node[inner sep = 0] (TA){}; %Target
		\draw let 
			\p1 = (TA) in
			(CA) -- (\x1, \y1+0.33\boxsize); %Link
			
		%CnotB
		\node[draw, circle, fill=black, minimum width=0.14cm, inner sep=0pt, right= 151.1pt of Rph] (CB) {}; % Control
		\draw let  
    			\p1 = (CB) in
      			(\x1,\y1-\wirespace) circle (0.33\boxsize)
			node[inner sep = 0] (TB){}; %Target
		\draw let 
			\p1 = (TB) in
			(CB) -- (\x1, \y1-0.33\boxsize); %Link

		%beginning of C %420pt
		\node (debutC) at ([xshift=400pt, yshift=-0.5\wirespace]debut1){};
		
		%initial descriptor of C
		\node[above right = -5pt and 0pt of debutC] (qC0) {\scriptsize{$\bol s_{C}$}};

		%Control + 2
		\node[draw, rectangle, minimum width=\boxsize, minimum height=\boxsize, inner sep = 0, right=20pt of debutC] (U2){${+2}$};
		\node[draw, circle, fill=black, minimum width=0.14cm, inner sep=0pt]  (C2) at ([yshift=1.5\wirespace]U2){};% Control
		\draw (C2) -- (U2); %link
		
		%Control + 1
		\node[draw, rectangle, minimum width=\boxsize, minimum height=\boxsize, inner sep = 0, right=27pt of U2] (U1){${+1}$};
		\node[draw, circle, fill=black, minimum width=0.14cm, inner sep=0pt]  (C3) at ([yshift=-1.5\wirespace]U1){};% Control
		\draw (C3) -- (U1); %link

		%descriptor4
		\node[right = 10pt of qA0] (qA4) {\scriptsize{$\left (q_{Ax} \hspace{1pt},\, q_{Az}(-\s{\theta} q_{1z}q_{2x} + \cc{\theta} q_{1x})\right )$}};
		\node[right = 10pt of qB0] (qB4) {\scriptsize{$\left (q_{Bx} \hspace{1pt},\, q_{Bz}(-\s{\phi} q_{2x} + \cc{\phi} q_{2z}q_{1x})\right )$}};

		%descriptor 5
		\node[above right = -11pt and -1pt of U2] (qC5) {\scriptsize{$\bol s_{C}(5)$}};
		
		%descriptor 6
		\node[above right = -11pt and -1pt of U1] (qC6) {\scriptsize{$\bol s_{C}(6)$}};
		
		%end of wires
		\node[right = 370pt of debut1] (fin1) {};
		\node[right = 370pt of debut2] (fin2) {};
		\node[right = 220pt of debutA] (finA) {};
		\node[right = 220pt of debutB] (finB) {};
		\node[right = 112pt of debutC] (finC) {};
		
		%wires
		\draw (debut1) -- (H);
		\draw (H) -- (Rth);
		\draw (Rth) -- (fin1);
		\draw (debut2) -- (Rph);
		\draw (Rph) -- (fin2);
		\draw (debutA) -- (finA);
		\draw (debutB) -- (finB);
		\draw[line width=0.7mm] (debutC) -- (U2);
		\draw[line width=0.7mm] (U2) -- (U1);
		\draw[line width=0.7mm] (U1) -- (finC);
		
		%wire labels
		\node[below right = -6pt and -2pt of debut1] (w1) {\scriptsize{$\mathfrak Q_1$}};
		\node[below right = -6pt and -2pt of debut2] (w2) {\scriptsize{$\mathfrak Q_2$}};
		\node[below right = -6pt and -2pt of debutA] (wA) {\scriptsize{$\mathfrak Q_A$}};
		\node[below right = -6pt and -2pt of debutB] (wB) {\scriptsize{$\mathfrak Q_B$}};
		\node[below right = -6pt and -2pt of debutC] (wC) {\scriptsize{$\mathfrak S_C$}};
		
		 %time labels
		\node (marker0) at ([xshift=30pt, yshift=0pt]debut2){}; 
		\node (t0) at ([xshift=30pt, yshift=-2\wirespace]debut2) {\scriptsize{$t=0$}};
		\draw[dotted] (t0) -- (marker0);
		\node (marker3) at ([xshift=25pt, yshift=0pt]Rph){}; 
		\node (t3) at ([xshift=25pt, yshift=-2\wirespace]Rph) {\scriptsize{$t=3$}};
		\draw[dotted] (t3) -- (marker3);
		\node (marker4) at ([xshift=25pt, yshift=0pt]TB){}; 
		\node (t4) at ([xshift=25pt, yshift=-1\wirespace]TB) {\scriptsize{$t=4$}};
		\draw[dotted] (t4) -- (marker4);
		\node (marker6) at ([xshift=25pt, yshift=0pt]C3){}; 
		\node (t6) at ([xshift=25pt, yshift=-1\wirespace]C3) {\scriptsize{$t=6$}};
		\draw[dotted] (t6) -- (marker6);

	\end{tikzpicture}
	}
	%\vspace{10pt}
	\caption{
A Bell experiment.
Thanks to the locality of descriptors, the calculations are performed directly on the wires, with the help of equations~\eqref{eqH}, \eqref{eqR} and~\eqref{eqCnot}.
A descriptor with no time label indicates its initial form, which for qubits is set by equation~\eqref{eq:initialdescriptor}.
The notations~$\s\gamma$ and~$\cc\gamma$ stand for $\sin \gamma$ and~$\cos \gamma$.
}
	\label{fig:bell}
\end{figure}

\emph{Upon measuring her particle of the entangled pair, Alice smoothly and locally evolves into}
\bes \label{eq:alicefoliates}
\qbt{A}{4} = \underbrace{P_{+1}(q_{1z}(3)) \qbt{A}{3}}_{\qbt{A, +1}{4}} + \underbrace{P_{-1}(q_{1z}(3)) \Uf^{\dagger}_\notgate\!\left(\qbt{A}{3}\right)\, \qbt{A}{3}\, \Uf_\notgate\!\left(\qbt{A}{3}\right)}_{\qbt{A, -1}{4}}
\ees
Because Alice is entangled,~$\bol q_A(4)$ admits no sharp observable.
Indeed, the expectation values of~$q_{Ax}(4)$, $q_{Az}(4)$ and \mbox{$q_{Ay}(4) = i\, q_{Ax}(4)\, q_{Az}(4)$} are all zero, which is incompatible with sharpness.
Yet as \mbox{$q_{Az} \equiv q_{Az}(0)$} is sharp with value~$1$, within each instance \mbox{$\bol q_{A,\,\pm 1}(4) = P_{\pm1}(q_{1z}(3))\,(q_{Ax},\, \pm q_{Az})$}, a $\pm 1$ $z$-outcome is indicated; each with measure~$1/2$ (see Appendix~\ref{sec:measures}).
At spacelike separation, Bob similarly foliates as $\bol q_B(4) = \bol q_{B,+1}(4) + \bol q_{B,-1}(4)$ when---and only when---\emph{he} measures.
Unlike relative states of the Schrödinger picture, relative descriptors are entirely local entities; they only concern individual localised systems.
Therefore, Alice's and Bob's foliations are independent of each other.\footnote{This solves a problem that Schrödinger-picture Everettian quantum mechanics does not solve, which is best expressed by Vaidman~\cite{VaidmanMWI}: `Although the [Many Worlds Interpretation] removes the most bothersome aspect of nonlocality, action at a distance, the other aspect of quantum nonlocality, the nonseparability of remote objects manifested in entanglement, is still there. A ``world'' is a nonlocal concept. This explains why we observe nonlocal correlations in a particular world'.}

Crucially, \emph{to confirm the violation of Bell inequalities, Alice and Bob must further interact to produce a record of the joint outcomes.}
In the single-world orthodoxy, bringing systems together that are each in a definite state merely amounts to convenience for comparison.
But in unitary quantum theory, that comparison needs to be analysed within the theory and, as shall be seen, permits nontrivial arrangements of the outcomes' measures.
Without loss of generality, Charlie first processes Alice's communication, so between time~$4$ and~$5$, he foliates with respect to~$q_{Az}(4)$, in a way that mirrors the foliations previously encountered,
$\bol s_C(5) = \bol s_{C,+1}(5) + \bol s_{C,-1}(5)$,
with
\beas
\bol s_{C,+1}(5) &=& P_{+1}(q_{Az}(4))\, \bol s_C \\
\bol s_{C,-1}(5) &=& P_{-1}(q_{Az}(4))\, \Uf_{\plustwo}^\dagger(\bol s_C)\, \bol s_C\, \Uf_{\plustwo}(\bol s_C)\,.
\eeas
When Charlie processes Bob's communication, both~$\bol s_{C,+1}(5)$ and~$\bol s_{C,-1}(5)$ foliate again:
\bea \label{eq:merge}
\bol s_C(6)
&=& \Uf_{\cplusone}^\dagger\!\left(\bol q_B(5),\, \bol s_{C,+1}(5)\right)\, \bol s_{C,+1}(5)\, \Uf_{\cplusone}\!\left(\bol q_B(5),\, \bol s_{C,+1}(5)\right) \nonumber\\
&& +\, \Uf_{\cplusone}^\dagger\!\left(\bol q_B(5),\, \bol s_{C,-1}(5)\right)\, \bol s_{C,-1}(5)\, \Uf_{\cplusone}\!\left(\bol q_B(5),\, \bol s_{C,-1}(5)\right) \nonumber\\[5pt]
&=& P_{+1}(q_{Bz}(5))\, P_{+1}(q_{Az}(4))\, \bol s_C \\
&& +\, P_{-1}(q_{Bz}(5))\, P_{+1}(q_{Az}(4))\, \Uf_{\plusone}^\dagger(\bol s_C)\, \bol s_C\, \Uf_{\plusone}(\bol s_C) \nonumber\\
&& +\, P_{+1}(q_{Bz}(5))\, P_{-1}(q_{Az}(4))\, \Uf_{\plustwo}^\dagger(\bol s_C)\, \bol s_C\, \Uf_{\plustwo}(\bol s_C) \nonumber\\
&& +\, P_{-1}(q_{Bz}(5))\, P_{-1}(q_{Az}(4))\, \Uf_{\plusthree}^\dagger(\bol s_C)\, \bol s_C\, \Uf_{\plusthree}(\bol s_C)\,.\nonumber
\eea
\emph{The record arises from the two local worlds of Alice, and those of Bob, and foliates into four non-interacting instances that respectively indicate $00$, $01$, $10$ and $11$}, in binary.
As is calculated in Appendix~\ref{sec:measures}, the respective measures of those instances are
\be \label{eqmes}
\frac{1}{2}\left(
\cos^2\frac{\theta - \phi}{2},\;
\sin^2\frac{\theta - \phi}{2},\;
\sin^2\frac{\theta - \phi}{2},\;
\cos^2\frac{\theta - \phi}{2}
\right)\,.
\ee
If the angles are optimally chosen, a~$\cos^2(\pi/8)$ win rate in a CHSH test is obtained (see Appendix~\ref{sec:bell}).

\section{Relativity}

Consider the relativistic twin paradox involving Alice, who has been staying on Earth, and her brother Bob, who has been travelling at high velocity and is currently far away.
Should they eventually reunite on Earth, which twin shall be the younger?
Invoking that the traveller is younger to \emph{de facto} declare Bob younger is generically mistaken, for there is no fact of the matter concerning age relations at spacelike separation.
Indeed, different unfoldings of the story lead to different relations.
In the most common continuation, Bob returns to Earth and finds himself younger than his stationary sister.
Yet, imagine a twist where, during Bob's journey, Alice embarks on an even faster and more extensive voyage.
When she returns to Earth alongside Bob, she is younger than her sibling.

Similarly, in unitarity of quantum theory, there is no fact of the matter about the relationship between spacelike separated worlds because local worlds can, in principle, be altered in a way that affects their future matching.
Like with the travelling twins, different continuations lead to different relationships, which are only measured once records about them are generated.
For instance, suppose Alice receives input~`$0$' for the CHSH test, and, abiding by the usual protocol (exposed in Appendix~\ref{sec:bell}), performs no rotation, i.e.~$\theta_0 = 0$.
Consider now the Alice who measured output~`$0$' (or $+1$ of the $z$ observable).
What can she say about the Bob that she will meet?
After Bob has followed the protocol and measured, it seems absolute that Alice will meet, within sub-measure~$\cos^2(\pi/8)$, a Bob who has also seen output~`$0$'.
But that is the same mistake that assumes the common continuation.
Since no measurement is definitive, Bob—or rather Wigner~\cite{wigner1961remarks}, who has coherent control over Bob and his particle—can in principle undo Bob's measurement, alter the particle's rotation with~$\Ry{\pi - \phi}$, re-measure, and proceed towards Alice to compare results.
The Alice who observed~`$0$' then meets with the Bob who observed~`$1$' with measure unity.
This practically hard, yet theoretically possible, continuation contradicts the absoluteness of the relationship between worlds before systems composing them interact.
\emph{Bell violations do not occur at spacelike separation.}

\section{Classicality}

The attempted completion of quantum theory by hidden variables is an aspiration to explain the quantum domain with an underlying classical reality.
Abiding by the probability calculus that underpins Shannon's theory~\cite{shannon}, hidden variables amount to pieces of genuinely classical information.
They are the supposed information missing to determine measurement outcomes. 
Bell's theorem shows that, should they exist, such pieces of classical information cannot be constrained by the causal structure of spacetime, so only conspicuous and problematic mechanisms can save the program (e.g. non-locality, retrocausality or superdeterminism).
\emph{But genuinely classical information does not exist}.

With the unrivalled success of quantum theory, the tables must be turned, and it is the classical realm that demands an underlying quantum explanation.
That is the role of decoherence theory~\cite{zeh1970interpretation, zurek1981pointer, zurek1982environment, gell1993classical, joos2003decoherence}, which studies how interactions involving also the environment give rise to quasi-classical domains.

\begin{figure}
	\centering
	\vspace{-15pt}
	\resizebox{\linewidth}{!}{
	\begin{tikzpicture} [scale=1,  line width=0.75] 
		%\centering
		
		%beginning of wire
		\node (debut1) at (0,0){};
		\node (debutE) at ([xshift=12pt, yshift=\wirespace+\wiresmall]debut1){};
		\node (debut2) at ([xshift=0pt, yshift=-3\wiresmall]debut1){};

		%initial descriptors
		\node[above right = -5pt and 0pt of debut1] (q13) {\scriptsize{$(q_{1x}(3), q_{1z}(3))$}};
		\node[above right = -5pt and 0pt of debut2] (q23) {\scriptsize{$(q_{2x}(3), q_{2z}(3))$}};
		\node[above right = -5pt and 0pt of debutE] (qE0) {\scriptsize{$(\bar q_{Ex}, \bar q_{Ez})$}};
			
		%CnotE
		\node[draw, circle, fill=black, minimum width=0.14cm, inner sep=0pt, right= 61pt of debut1] (CE) {}; % Control
		\draw let  
    			\p1 = (CE) in
      			(\x1,\y1+\wirespace+\wiresmall) circle (0.33\boxsize)
			node[inner sep = 0] (TE){}; %Target
		\draw let 
			\p1 = (TE) in
			(CE) -- (\x1, \y1+0.33\boxsize); %Link
			
		%descriptor1 after E
		\node[right = -2pt of q13] (q13h) {\scriptsize{$\left (q_{1x}(3) \bar q_{Ex} \hspace{1pt},\, q_{1z}(3) \right )$}};

		%CnotA
		\node[draw, circle, fill=black, minimum width=0.14cm, inner sep=0pt, right= 74pt of CE] (CA) {}; % Control
		\draw let  
    			\p1 = (CA) in
      			(\x1,\y1+\wirespace) circle (0.33\boxsize)
			node[inner sep = 0] (TA){}; %Target
		\draw let 
			\p1 = (TA) in
			(CA) -- (\x1, \y1+0.33\boxsize); %Link
			
		%Wire A
		\node (debutA) at ([xshift=89pt, yshift=\wirespace]debut1){};
		\node[above right = -5pt and 0pt of debutA] (qA0) {\scriptsize{$(q_{Ax} \,, \, q_{Az})$}};
	
		%desA after Cnot 
		\node[right = 8pt of qA0] (qA4) {\scriptsize{$\left (q_{Ax} \hspace{1pt},\, q_{Az} q_{1z}(3)\right )$}};
		%(-\s{\theta} q_{1z}q_{2x} + \cc{\theta} q_{1x})
			
		%CnotA'
		\node[draw, circle, fill=black, minimum width=0.14cm, inner sep=0pt, right= 72pt of TA] (CA') {}; % Control
		\draw let  
    			\p1 = (CA') in
      			(\x1,\y1-\wiresmall) circle (0.33\boxsize)
			node[inner sep = 0] (TA'){}; %Target
		\draw let 
			\p1 = (TA') in
			(CA') -- (\x1, \y1-0.33\boxsize); %Link
		
		% Wire A'	
		\node (debutA') at ([xshift=-62pt]TA'){};
		\node[above right = -5pt and 0pt of debutA'] (qA'0) {\scriptsize{$(q_{A'x} \,, \, q_{A'z})$}};
		%desA' after Cnot 
		\node[right = 30pt of qA'0] (qA'4) {\scriptsize{$\left (q_{A'x} \hspace{1pt},\, q_{Az} q_{A'z}q_{1z}(3)\right )$}};
		
		%CnotA''
		\node[draw, circle, fill=black, minimum width=0.14cm, inner sep=0pt, right= 20pt of TA'] (CA'') {}; % Control
		\draw let  
    			\p1 = (CA'') in
      			(\x1,\y1-\wiresmall) circle (0.33\boxsize)
			node[inner sep = 0] (TA''){}; %Target
		\draw let 
			\p1 = (TA'') in
			(CA'') -- (\x1, \y1-0.33\boxsize); %Link
			
		% Wire A''
		%\node (debutA'') at ([yshift=-\wiresmall]debutA'){};
		\node (debutA'') at ([xshift=-67pt]TA''){};
		\node[above right = -5pt and 0pt of debutA''] (qA''0) {\scriptsize{$(q_{A''x} \,, \, q_{A''z})$}};
		
		%desA'' after Cnot 
		\node[right = 30pt of qA''0] (qA''4) {\scriptsize{$\left (q_{A''x} \hspace{1pt},\, q_{Az} q_{A'z} q_{A''z}q_{1z}(3)\right )$}};
		
		% Fleche descritpeurs
		\node[inner sep = 0] (debutfleche) at ($(TA')!0.5!(CA'')$) {};
		\draw[>={Latex[length=1mm]}, <->, line width=0.1mm] (debutfleche)  to [out=60, in=180] (qA'4.west);

		% Wire C
		\node (debutC) at ([xshift= 50pt, yshift=-\wiresmall]debutA''){};
		
		\node[above right = -5pt and 0pt of debutC] (qC0) {\scriptsize{$\bol s_{C}$}};
		
		%Control + 2
		\node[draw, circle, fill=black, minimum width=0.14cm, inner sep=0pt]  (C2) at ([xshift=20pt]TA''){};% Control
		\node[draw, rectangle, minimum width=\boxsize, minimum height=\boxsize, inner sep = 0] (U2) at ([yshift=-\wiresmall]C2){${+2}$};
		\draw (C2) -- (U2); %link

		% Fleche descritpeurs
		\node[inner sep = 0] (debutfleche) at ($(TA'')!0.5!(C2)$) {};
		\draw[>={Latex[length=1mm]}, <->, line width=0.1mm] (debutfleche)  to [out=60, in=180] (qA''4.west);
		
		%Control + 1
		\node[draw, rectangle, minimum width=\boxsize, minimum height=\boxsize, inner sep = 0, right=9.5pt of U2] (U1){${+1}$};
		\node[draw, circle, fill=black, minimum width=0.14cm, inner sep=0pt]  (C3) at ([yshift=-\wirespace]U1){};% Control
		\draw (C3) -- (U1); %link

		% Wire B''
		\node (debutB'') at ([xshift=-123pt]C3){$\dots$};
		%\node (debutB'') at ([yshift=-2\wirespace]debutA''){};
		%desc
		\node[above right = -5pt and 0pt of debutB''] (qb'') {\scriptsize{$\left (q_{B''x} \hspace{1pt},\, q_{Bz} q_{B'z} q_{B''z}q_{2z}(3)\right )$}};
	
%		%descriptor 5
%		\node[above right = -11pt and 0pt of U2] (qC5) {\scriptsize{$q_C(5)$}};	
		%descriptor 6
		\node[above right = -11pt and 0pt of U1] (qC6) {\scriptsize{$\bol s'_{C}(6)$}};
		
		% 'Decoherence'
		\node[below left = 0pt and 15 pt of CE] (brace1) {}; 
		\node[below right = 0pt and 15 pt of CE] (brace2) {};
		\draw[decorate,decoration={brace,amplitude=5pt,mirror}] 
  (brace1) -- (brace2) node[midway,below,yshift=-5pt] {\scriptsize{Decoherence}};
  
 		% 'Chain reaction
		\node[above right = 5pt and 25 pt of CA'] (brace3) {}; 
		\node[above right = 0pt and 28 pt of CA''] (brace4) {};
		\draw[decorate,decoration={brace,amplitude=5pt}] 
  (brace3) -- (brace4) node[midway,xshift=36pt, yshift=8pt] {\scriptsize{Chain reaction}};
		
		%end of wires
		\node[right = 150pt of debut1] (fin1) {};
		\node[right = 62pt of debut2] (fin2) {$\dots$};
		\node[right = 150pt of debutA] (finA) {};
		\node[right = 80pt of debutE] (finE) {};
		\node[right = 172pt of debutA'] (finA') {};
		\node[right = 198pt of debutA''] (finA'') {};
		\node[right = 140pt of debutB''] (finB'') {};
		\node[right = 148pt of debutC] (finC) {};
		
		%wires
		\draw (debut1) -- (fin1);
		\draw (debut2) -- (fin2);
		\draw (debutE) -- (finE);
		\draw (debutA) -- (finA);
		\draw (debutA') -- (finA');
		\draw (debutA'') -- (finA'');
		\draw (debutB'') -- (finB'');
		\draw[line width=0.7mm] (debutC) -- (U2);
		\draw[line width=0.7mm] (U2) -- (U1);
		\draw[line width=0.7mm] (U1) -- (finC);
		
		%wire labels
		\node[below right = -6pt and -2pt of debut1] (w1) {\scriptsize{$\mathfrak Q_1$}};
		\node[below right = -6pt and -2pt of debut2] (w2) {\scriptsize{$\mathfrak Q_2$}};
		\node[below right = -6pt and -2pt of debutE] (wE) {\scriptsize{$\mathfrak Q_E$}};
		\node[below right = -6pt and -2pt of debutA] (wA) {\scriptsize{$\mathfrak Q_A$}};
		\node[below right = -6pt and -2pt of debutA'] (wA') {\scriptsize{$\mathfrak Q_{A'}$}};
		\node[below right = -6pt and -2pt of debutA''] (wA'') {\scriptsize{$\mathfrak Q_{A''}$}};
		\node[below right = -6pt and -2pt of debutC] (wC) {\scriptsize{$\mathfrak S_C$}};
		\node[below right = -6pt and -2pt of debutB''] (wB'') {\scriptsize{$\mathfrak Q_{B''}$}};
		
		 %time labels
		\node (marker3) at ([xshift=30pt, yshift=0pt]debut2){}; 
		\node (t3) at ([xshift=30pt, yshift=-0.8\wirespace]debut2) {\scriptsize{$t=3$}};
		\node (marker3') at ([yshift=0.5\wirespace]marker3){};
		\node (marker3'') at ([yshift=1.7\wirespace]marker3'){};
		\draw[dotted] (t3) -- (marker3);
		\draw[dotted] (marker3') -- (marker3'');
%		\node (marker3) at ([xshift=25pt, yshift=0pt]Rph){}; 
%		\node (t3) at ([xshift=25pt, yshift=-2\wirespace-10pt]Rph) {\scriptsize{$t=3$}};
%		\draw[dotted] (t3) -- (marker3);
%		\node (marker4) at ([xshift=25pt, yshift=0pt]TB){}; 
%		\node (t4) at ([xshift=25pt, yshift=-1\wirespace-10pt]TB) {\scriptsize{$t=4$}};
%		\draw[dotted] (t4) -- (marker4);
%		\node (marker6) at ([xshift=25pt, yshift=0pt]C3){}; 
%		\node (t6) at ([xshift=25pt, yshift=-1\wirespace-10pt]C3) {\scriptsize{$t=6$}};
%		\draw[dotted] (t6) -- (marker6);

	\end{tikzpicture}
	}
	\vspace{-20pt}
	\caption{
The Bell experiment is enhanced with considerations of classicality.
First, after~$\mathfrak Q_1$'s rotation, but before Alice's measurement, a decoherent interaction involves an environmental system, which contains at least the logical space of a qubit.
Unlike the initialized qubits, the environment's descriptor~$(\bar q_{Ex}, \bar q_{Ez})$ is given by some generic representation of the Pauli algebra, which generically differ from that of equation~\eqref{eq:initialdescriptor}.
Second, a chain reaction through properly initialized systems~$\mathfrak Q_{A'}$, $\mathfrak Q_{A''}$—and generically many more—models classical communication.
The nontrivial arrangement of~$\mathfrak S_C$'s relative descriptors is unhindered by these considerations of classicality.
}
	\label{fig:dec}
\end{figure}

Here I shall show that the processes involving~$\mathfrak Q_A$, $\mathfrak Q_B$ and~$\mathfrak S_C$ are classical, according to explanations of what ‘classical’ can mean \emph{within} quantum theory.
First, as illustrated in Figure~\ref{fig:dec}, after the rotation of Particle~1, a nearby environment can be modelled to interact with~$\mathfrak Q_1$ in a way that stabilizes the $z$ observable to be measured.
The effect is that the unstructured~$\bar q_{Ex}$ is copied onto the $x$ component of~$q_{1}(3)$, obfuscating~$q_{1x}(3)$ and thus decohering~$\mathfrak Q_1$ into its $z$ basis.
Yet~$\bar q_{Ex}$ is \emph{not} copied onto the following systems.
Therefore, decoherence as such neither prevents the mergings of Alice’s and Bob’s worlds portrayed in equation~\eqref{eq:merge} nor, therefore, the ability to violate Bell’s inequalities.
Decoherence may also affect~$\mathfrak Q_A$, $\mathfrak Q_{A'}$, $\mathfrak Q_{A''}$, analogous systems on Bob’s side, or, with appropriate generalizations,~$\mathfrak S_C$—with no effects on the nontrivial arrangement of~$\mathfrak S_C$’s relative descriptors.
Such robustness to decoherence is a distinguishing property of classical processes.

Second, the realistic physical processes that amount to measuring and eventually communicating an experiment’s result involve more than a single binary system.
In fact, in the kind of communication we call ‘classical’, a precise quantum system is \emph{not} sent from one location to another;
rather, the information is transmitted through a chain reaction in a collection of quantum systems, like~$\mathfrak Q_{A'}$, $\mathfrak Q_{A''}$ and generically many more.
The resulting descriptor~$\bol s'_{C}(6)$ admits, like in equation~\eqref{eq:merge}, a foliation into four entities, which are only altered insofar as the arguments in~$P_{\pm 1}(\cdot)$ acquire extra factors of~$q_{A'z} q_{A''z}$ and~$q_{B'z} q_{B''z}$.
As these operators have eigenvalue~$1$ with respect to the Heisenberg state, the measures in equation~\eqref{eqmes} are invariant, and so the foliation of~$\mathfrak S_C$ is not changed in any observable way.
Intermediary systems analogous to~$\mathfrak Q_{A'}$ and~$\mathfrak Q_{A''}$ can also represent a mediation between Particle~1 and Alice; again, without consequences on the foliation of~$\mathfrak S_C$.
This marks a second distinctively classical feature: \emph{chain reactions enable what appears as classical communication}, which—crucially—does not preclude the violation of Bell inequalities.

Moreover, $\mathfrak Q_A$, $\mathfrak Q_{A'}$, $\mathfrak Q_{A''}$, etc., can be understood as instantiating parts of a macroscopic Alice and her surroundings.
In this light, although the foliation of~$\mathfrak Q_A$ presented in section~\ref{sec:exp} is one~$\cnot$ away from recombining, this fragility is removed when Alice consists of many more degrees of freedom.
As the unsharp observable~$q_{1z}(3)$ gets copied through the systems, they foliate.
Therefore, \emph{everything that suitably interacts with the Alices foliates in turn, creating worlds which, for all practical purposes, are distinct and autonomous.}

\section{Conclusion}

I showed that a fully local account of Bell violations is precisely and formally instantiated in unitary quantum theory. In doing so, I explored how sets of local worlds, generated by remote foliations, interact in their future lightcone. When entanglement relates the unsharp observables responsible for the foliations, a non-trivial arrangement of the worlds’ measures arises as they later interact. This kind of interference persists even when systems decohere. Therefore, joining lists to compare statistics, a seemingly mundane and classical operation, is not trivial in the multiverse. This non-trivial arrangement follows from quantum theory alone, enabling Bell correlations while still preserving locality. For those of us who regard the unitary quantum theory as universally valid and, in line with scientific realism, assign meaning to its representation in the Heisenberg picture, Bell’s conundrum is solved.

\subsection*{Acknowledgements}

I am grateful to David Deutsch and Samuel Kuypers for stimulating discussions and feedback on earlier versions of this article.
I am also grateful to Gilles Brassard, Andrea Di Biagio, Christopher Timpson, Luuk van den Berg and Xavier Coiteux-Roy for insightful comments, and I thank Xavier as well for providing the diagrams of Alice and Bob that I used in constructing figure~\ref{fig:diag}.
I wish to thank Claude Crépeau and Stefan Wolf for their invaluable support.
Finally, I also wish to thank the Conjecture Institute for welcoming me as a fellow and for its intellectual support, particularly through the careful feedback of Logan Chipkin.

This work was supported in parts by the Fonds de recherche du Québec -- Nature et technologie, the Swiss National Science Foundation, the Hasler Foundation and the Mitacs Elevate postdoctoral fellowship in partnership with Bbox Digital.

\appendix

\section{Bell's Theorem as a Game} \label{sec:bell}

A simple presentation of Bell's theorem is given by the \emph{CHSH game}~\cite{CHSH, cleve2004consequences, gisin2014quantum}.
The game is played cooperatively by two players, Alice and Bob, who know the game's rules and attempt to achieve the highest score.
The game occurs when the players are at spacelike separation, ensuring no communication takes place between them.
Each player is asked a binary question,~$0$ or~$1$, and produces a binary answer,~$0$ or~$1$.
They win the game if the product of the questions equals the parity of the answers.
In other words, if at least one question is~$0$, Alice and Bob must answer the same bit; and if both questions are~$1$, they must answer different bits.
Questions and answers are also referred to as inputs and outputs, respectively.
Before playing the game, the players are colocated and can agree on a strategy.
They can share pieces of classical information or throw coins for common randomness if they deem it helpful.

If Alice opts for a deterministic strategy, her output~$a$ is one of the four functions $\{0,1\}\!\to\!\{0,1\}$ of her input~$x$, and similarly, a deterministic strategy for Bob consists of $b=b(y)$, for a total of $16$ joint deterministic strategies.
To avoid the contradicting bottom line, one (or three) of the following equations must not hold:
\begin{equation*}
\begin{array}{rcl}
a(0) \oplus b(0) &=& 0 \\[2pt]
a(0) \oplus b(1) &=& 0 \\[2pt]
a(1) \oplus b(0) &=& 0 \\[2pt]
a(1) \oplus b(1) &=& 1 \\[2pt]
\cline{1-3}\\[-10pt]
0 &=& 1\,.
\end{array}
\end{equation*}
This means that deterministic strategies can, at best, win on~$3$ out of the~$4$ possible input pairs.
Assuming that the questions are independent of any prior knowledge of Alice and Bob, exchanging information or shared randomness cannot improve their cause, so their best expected win rate is~$3/4$.

\subsubsection*{Sharing Entanglement and the Quantum Strategy} \label{Bellwin}

\noindent Unless the players share entangled particles.
Let Particles~1 and~2 be in the state $\ket{\Phi^+} = (\ket{00} + \ket{11})/\sqrt 2$.
To beat the~$3/4$ win rate at the CHSH game, Alice and Bob can perform a Bloch-sphere rotation that depends on their input, as prescribed in Table~\ref{tab:angles}, before measuring their particle and returning the outcome as output.

\begin{table}[h]
\centering
\begin{tabular}{ccc}
Input & Alice's Rotation ($\theta$) & Bob's Rotation ($\phi$) \\
\hline
0 & $\theta_0 = 0$ & $\phi_0 = \frac{\pi}{4}\hphantom{-}$ \\
1 & $\theta_1 = \frac{\pi}{2}$ & $\phi_1 = -\frac{\pi}{4}$ \\
\end{tabular}
\caption{The quantum strategy with a shared~$\ket{\Phi^+}$.}
\label{tab:angles}
\end{table}

The angles are such that if at least one input is~`$0$', they differ by~$\pi/4$, and if both inputs are~`$1$', by~$3\pi/4$.
The measurement's probability distribution, given in Table~\ref{tab:stats}, yields an expected win rate of~$\cos^2(\pi/8)$.
Larger than~$3/4$.

\begin{table}[h]
\centering
\begin{tabular}{c c c c c c}
Input & \multicolumn{5}{c}{Distribution of Outcomes} \\
Pair & & $00$ & $01$ & $10$ & $11$ \\[1pt]
\hline
\vphantom{$\int^A$}$(0,0)$ & &
$\cos^2(\pi/8)/2$ &
$\sin^2(\pi/8)/2$ &
$\sin^2(\pi/8)/2$ &
$\cos^2(\pi/8)/2$
\\[2pt]
$(0,1)$ & &
$\cos^2(\pi/8)/2$ &
$\sin^2(\pi/8)/2$ &
$\sin^2(\pi/8)/2$ &
$\cos^2(\pi/8)/2$
\\[2pt]
$(1,0)$ & &
$\cos^2(\pi/8)/2$ &
$\sin^2(\pi/8)/2$ &
$\sin^2(\pi/8)/2$ &
$\cos^2(\pi/8)/2$
\\[2pt]
$(1,1)$ & &
$\sin^2(\pi/8)/2$ &
$\cos^2(\pi/8)/2$ &
$\cos^2(\pi/8)/2$ &
$\sin^2(\pi/8)/2$
\end{tabular}
\caption{Distribution of the quantum strategy for each input pair.}
\label{tab:stats}
\end{table}

\subsubsection*{From the Game Back to Physics}

How do entangled particles work?
The lesson of the CHSH game is that it provides a generic way by which they \emph{cannot} work:
The particles' apparently random outcomes cannot be determined by underlying parameters—hidden variables—shared in their common past and constrained by the causal structure of spacetime.
Many conclusions drawn from Bell's theorem are attempts to save the hidden variable program by confronting the causal structure.
Some turn to non-local actions between particles that can entail a secret coordination of the particles' response across space.
Others advocate superdeterminism, which, viewed in the CHSH game, amounts to positing that the particles knew the questions, so the information initially shared is unusually helpful in order to achieve a win rate better than~$3/4$.

\section{Calculating Measures}\label{sec:measures}

A few more properties of descriptors are provided, and the calculations of measures of~$\bol q_A(4)$ and~$\bol s_{C}(6)$ are explained.

\sloppy
Let~$G$ be the matrix representation of a gate acting nontrivially only on systems labelled by~$\mathcal J$.
Its functional representation~$\Uf_G$ satisfies the defining equation 
\be\label{eq:fform}
\Uf_{G}\left(\{\bol q_j(0)\}_{j \in \mathcal J}\right)=G \,,
\ee
which is guaranteed to exist by the generative ability of the components of $\{ \bol q_j(0)\}_{j \in \mathcal J}$.
For instance, if the gate affects qubits~$1$ and~$2$, 
then the matrix~$G$ can be expressed as polynomial in the matrices~$q_{1x}(0)$, $q_{1z}(0)$, $q_{2x}(0)$, $q_{2z}(0)$, and~$\Uf_{G}$ is one such polynomial.

\sloppy
Descriptors evolve as operators do, namely,
\bes %\label{eq:evotot}
\bol q_i(t) = U^{\dagger} \bol q_i(0) U \,,
\ees
where~$U$ denotes the matrix form of the evolution of the whole network between time~$0$ and time~$t$.
This more familiar evolution is equivalent to the step-by-step evolution of equation~\eqref{eq:step}.
Indeed let~$V$ and~$G$ be the matrices of the dynamical operators occurring between time~$0$ and~$t$ and between time~$t$ and~$t+1$, respectively.
Then,
\beas
%\bol q_i(t) %&=& U^\dagger \bol q_i(0) U \\
 V^\dagger G^\dagger  \bol q_i(0) G V 
&=& V^\dagger \Uf^\dagger_{G}\left(\{\bol q_j(0)\}_{j\in \mathcal J}\right) V\, V^\dagger \bol q_i(0)V\, V^\dagger \Uf_{G}\left(\{\bol q_j(0)\}_{j\in \mathcal J}\right) V \\
&=& \Uf^\dagger_{G}\left ( \{V^\dagger \bol q_j(0) V \}_{j\in \mathcal J} \right ) \, V^\dagger \bol q_i(0) V \, \Uf_{G}\left ( \{V^\dagger \bol q_j(0) V \}_{j\in \mathcal J} \right )\\
&=& \Uf^\dagger_{G}\left(\{ \bol q_j(t-1) \}_{j\in \mathcal J} \right) \, \bol q_i(t-1) \, \Uf_{G}\left(\{ \bol q_j(t-1) \}_{j\in \mathcal J} \right) \,.
\eeas
The second equality holds because in each term of the polynomial $\Uf_{G}\left ( \{V^\dagger \bol q_j(0) V \}_{j \in \mathcal J}\right )$, products will have their inner~$V^\dagger$s and~$V$s cancelled, leaving only the outer ones, which can be factorised outside of the polynomial to retrieve the first line.

As is usual for qubits, the $+1$ and $-1$ eigenvalues of the $\hat z$ observable are identified respectively with the computational values~$0$ and~$1$.
The multiversal measure of Alice witnessing outcome $i \in \{0,1\}$ is given by the expectation of the observable
$
\ketbra ii^{\mathfrak Q_A} \otimes \one^{\overline{\mathfrak Q_A}} 
%=P_{+1}(q_{Az}(0))\,.
$.
Letting~$U$ denote the whole unitary operator between time $0$ and $4$, and recalling that the Heisenberg state is~$\ket 0^{\otimes n} \equiv \ket{\bol 0}$, this expectation is
\beas
\left \langle
 U^\dagger \left( \ketbra ii^{\mathfrak Q_A} \otimes \one^{\overline{\mathfrak Q_A}} \right) U 
\right \rangle 
&=& \langle U^\dagger P_{(-1)^i}(q_{Az}(0)) U \rangle \\
&=& \langle  P_{(-1)^i}(q_{Az}(4)) \rangle \\
&=& \frac 12 + (-1)^i \frac{\bra{\bol 0}    q_{Az}(-\sin{\theta} q_{1z}q_{2x} + \cos{\theta} q_{1x})    \ket{\bol 0}}{2}\\
&=& \frac 12 \,.
\eeas

For $i$, $j \in \{0,1\}$, let~$\ket{ij}$ denote in binary the eigenstates of the computational observable of $\mathfrak S_C$.
The measure of Charlie witnessing $ij$ is given by the expectation of the observables $\ketbra{ij}{ij}^{\mathfrak{S}_C} \otimes \one^{\overline{\mathfrak{S}_C}}$, which can be expressed as a function~$f_{ij}(\bol s_{C}(0))$ of~$\mathfrak{S}_C$'s initial descriptor.
Without even knowing the form of~$ \bol s_{C}(0)$, 
its generative abilities guarantee that one can construct from it any functional form of a gate concerning $\mathfrak S_C$, as well as any observables pertaining to $\mathfrak S_C$.
At time $6$, the observables can be expressed with the help of equation~\eqref{eq:merge} as
\beas
f_{ij}(\bol s_{C} (6)) &=& P_{+1}(q_{Bz}(5)) \, P_{+1}(q_{Az}(4)) \, f_{ij}(\bol s_{C}) \\
&& \,+\,P_{-1}(q_{Bz}(5)) \, P_{+1}(q_{Az}(4)) \, \Uf_{\plusone}^\dagger(\bol s_{C})\, f_{ij}(\bol s_{C}) \, \Uf_{\plusone}(\bol s_{C})  \nonumber\\
&& \, + \, P_{+1}(q_{Bz}(5)) \, P_{-1}(q_{Az}(4)) \, \Uf_{\plustwo}^\dagger(\bol s_{C}) \, f_{ij}(\bol s_{C}) \, \Uf_{\plustwo}(\bol s_{C}) \nonumber\\
&& \, + \, P_{-1}(q_{Bz}(5)) \, P_{-1}(q_{Az}(4)) \, \Uf_{\plusthree}^\dagger(\bol s_{C})  \, f_{ij}(\bol s_{C}) \, \Uf_{\plusthree}(\bol s_{C})\,,
\eeas
where again~$\bol s_C$ without label refers to its initial form, which in this case is the same at times~$0$ and~$4$.
Using the defining equation~\eqref{eq:fform}, $\Uf_{\plusk}(\bol s_{C})$ amounts to the gate `$\plusk$' in its matrix form. 
This is useful to compute $\bra{\bol 0} f_{ij}(\bol s_C (6)) \ket{\bol 0}$ from the right.
In each term, `$\plusk$' transforms the Heisenberg state to~${\ket k}^{\mathfrak S_C} \otimes {\ket{\bol 0}}^{\overline{\mathfrak S_C}}$, which cancels upon finding $f_{ij}(\bol s_{C})$ except for the term where $k=ij$ in binary. 
The operator $\Uf_{\plusk}^\dagger(\bol s_{C})$ transforms back $\ket{k}$ to $\ket 0$, leaving $\langle P_{(-1)^i}(\bol q_{Az}(4)) P_{(-1)^j}(\bol q_{Bz}(5)) \rangle$. This can then be calculated:
\beas
&& \left \langle P_{(-1)^i}(q_{Az}(-\s{\theta} q_{1z}q_{2x} + \cc{\theta} q_{1x})) 
P_{(-1)^j}(q_{Bz}(-\s{\phi} q_{2x} + \cc{\phi} q_{2z}q_{1x} ))  \right \rangle \\
&=& \frac 14 + (-1)^{i+j} \langle (q_{Az}(-\s{\theta} q_{1z}q_{2x} + \cc{\theta} q_{1x}))(q_{Bz}(-\s{\phi} q_{2x} + \cc{\phi} q_{2z}q_{1x} )) \rangle \\
&=& \frac 14 + (-1)^{i+j} \frac{(\s{\theta}\s{\phi}  + \cc{\theta}\cc{\phi})}{4}\\
&=& \frac 14 +(-1)^{i+j}  \frac{(\cos(\theta - \phi))}{4}\\
&=& \begin{cases}
 \frac 12  \cos^2 \left(\frac{\theta - \phi}{2}\right) & \text{if } i=j\\
 \frac 12  \sin^2 \left(\frac{\theta - \phi}{2}\right) & \text{if } i\neq j \,.
\end{cases}
\eeas

\end{document}